\documentclass[10pt,letterpaper]{article}
\usepackage[top=0.85in,left=1.5in,footskip=0.75in,marginparwidth=1.5in]{geometry}

\usepackage[utf8]{inputenc}

\usepackage{cite}

\usepackage{nameref,hyperref}

\usepackage[right]{lineno}

\usepackage{microtype}
\DisableLigatures[f]{encoding = *, family = * }

\raggedright
\usepackage{changepage}

\usepackage[aboveskip=1pt,labelfont=bf,labelsep=period,singlelinecheck=off]{caption}

\makeatletter
\renewcommand{\@biblabel}[1]{\quad#1.}
\makeatother

\usepackage{lastpage,fancyhdr,graphicx}
\usepackage{epstopdf}
\fancyheadoffset[L]{2.25in}
\fancyfootoffset[L]{2.25in}

\usepackage{color}

\definecolor{Gray}{gray}{.25}

\usepackage{graphicx}

\usepackage{sidecap}

\usepackage{wrapfig}
\usepackage[pscoord]{eso-pic}
\usepackage[fulladjust]{marginnote}

\usepackage{graphicx,dblfloatfix}
\usepackage{subfigure}
\usepackage{mathtools}
\usepackage{amsmath,amssymb,amsfonts}
\usepackage{algorithmic}
\usepackage{textcomp}
\usepackage{amsmath,epsfig}
\usepackage{float}
\usepackage[lined,ruled,linesnumbered]{algorithm2e}
\usepackage{algorithmic}
\usepackage{ulem}
\usepackage{placeins}
\usepackage[numbers]{natbib}
\usepackage{array}

\reversemarginpar

\begin{document}
\vspace*{0.35in}

\begin{flushleft}
{\Large
\textbf\newline{Cyber Physical Systems (CPS) Surveillance Using An Epidemic Model}
}
\newline
\\
Bagula, A.\textsuperscript{1,*},
Tuyishimire, E.\textsuperscript{1},
Ajayi, O.\textsuperscript{1}
\\
\bigskip
\bf{1} ISAT Laboratory, Department of Computer Science, University of the Western Cape, South Africa
\\
\bigskip
* abagula@uwc.ac.za

\end{flushleft}

\section*{Abstract}
Vast investments have recently been made worldwide in developing the Cyber-Physical System (CPS) technology with the expectations of improving economical and societal structures. However, great care must be paid to the CPS' complexity, the impact of 
emerging IoT (Internet of Things) protocols on the CPS infrastructure as well as the impact of information dissemination by these protocols on the safety of these infrastructures.  This paper addresses the issue of CPS safety by proposing and evaluating the 
performance of a CPS management framework and the analysis of the dynamics of the underlining IoT network in the cyber-space. The main contributions of this paper are in threefold. Firstly, a new CPS framework is proposed; that: 1) builds around a layered architecture to compartmentalise the CPS functionalities into different modules for efficiency and scalability and 2) uses an inner feedback loop for the efficient management of CPS infrastructure. Secondly, building upon this framework, a novel diffusion model that uses the epidemic (interference) sets to produce accurate diffusion patterns across the CPS'  IoT subsystem  is proposed. Finally, the proposed diffusion model is numerically analysed to show how it can be used to achieve efficient CPS surveillance in order to trigger reconfiguration to re-optimise the CPS when it is under stress.  in IoT settings.  The numerical analysis of the diffusion model shows that interference propagates in pairwise disjoint sets, with IoT nodes migrating from ``susceptible'' to ``attacked'' statuses and finally reaching the ``removed'' state at a predictable time. Deployment considerations on some of the current social and public networks are also considered..


\section*{Introduction}
In recent times, vast investments have been made globally in the development of Cyber Physical Systems (CPS) technologies. It is expected that CPS would pave the way for solutions to key economic and societal challenges such as coping with an ageing population, addressing climate change, improving issues of health and public safety, supporting the switch to renewable energy, planning for megacities, tackling limited resources, achieving sustainability and globalisation as well as proffering solutions to mobility challenges. Similarly, Internet-of-Things (IoT) principles are finding their way into the next generation CPS  to enable extended interactive functionality between the physical  and virtual environments. CPS are a new generation of systems that play a key role in interconnecting the physical and virtual worlds. This is achieved by integrating computing and communication capabilities with the dynamics of physical and engineered systems; and is expected to provide different ways of interacting and manipulating physical systems through seamless network connectivity and refined user control over the actuation side.  

\subsection*{The Impact of the IoT subsystem on the CPS}

CPS and IoT technologies are currently enjoying tremendous attention globally. It is believed that these technologies can improve economical and societal structures as well as extend the interactive functionality between real and virtual environments. To this end, large amount of funds and efforts are being put into them. However, great considerations must be paid to the impacts of the complex emerging IoT platforms and standards on the CPS infrastructure. These impacts can be viewed from three dimensions viz: 

\begin{itemize}
\item
{\bf The Impact of Complexity on CPS.} By enabling the information to be collected and communicated among everybody, everything and anything, the IoT-aware CPS will usher in a new era where cyberspace, physical space, human knowledge and social activities are integrated into a universal platform. With these various components synchronised, we gain the ability to monitor the real world in ways that we could not fathom possible without the IoT. However, such benefits come at the expense of complex CPS platforms with heterogeneous components, multiple functionalities, rules and feedback loops. These could lead to new kinds of risks and vulnerabilities such as - rapid spread of hazards, faults or disturbances across entire systems of devices as a result of domino or cascading effects. Such disturbances which previously would have been localised to a single device, could evolving into large-scale, system-wide failure or disaster(s) if uncontained.  
\item
{\bf The Impact of Standards on CPS.} Current generation CPS are managed by Networked Control Systems (NCS), wherein physical processes are controlled by networks of sensors, actuators and controllers. These are often built around static topologies with pre-planned routing and scheduling mechanisms mandated by standard wireless protocols such the WirelessHART ~\cite{1b}. Though these standards provide real-time guarantees for delay-sensitive applications, they do not consider performance-related tasks such as management of resources at the physical layer, link layer scheduling of nodes and\textbackslash or end-to-end network layer routing of traffic flows. It is predicted that the next generation CPS will be built around performance aware NCS that rely on its IoT subsystem for the sensing and actuation on the environment being controlled.  These would take advantage of the IoT communication infrastructure and its lightweight protocols to support sensing and actuation in CPS. They would also provide varied services to users while meeting the requirements of high throughput, high reliability and energy efficiency while operating within bounded communication delays~\cite{qos,Multipath2}. 
\item
{\bf The Impact of Information Diffusion on CPS Safety.} The CPS of the future will be designed around a network infrastructure that interconnect islands of IoT networks, with information diffusion implemented on tree-like network topologies rooted at the gateway(s). The management of such forest of interconnected trees requires formal modelling and accurate performance analysis before and during deployment. Collection tree protocols~\cite{RPL,CTP,2} are rapidly gaining ground in the IoT field. They are protocols that rely on a spanning tree structure rooted at the sink of a sensor network to enable information diffusion from sensor nodes to the sink (which is usually connected to a gateway). The management of the information diffusion in collection tree based infrastructures is a challenging issue that needs to be addressed efficiently in order to avoid local disturbances escalating into global disasters.
\end{itemize}

\subsection*{Contributions and outline}

The discovery of information flow within the proposed framework and its impact on the network engineering process are two key issues that can positively affect the CPS efficiency. These issues can be described as follows:

{\bf Diffusion pattern discovery:} The diffusion of information in sensor networks usually follows a pattern that defines the network connectivity in terms of interactions between the nodes, the effect they might have on each others and the routing of information from nodes to the gateway. The discovery of such pattern and its underlying data structure is therefore a vital process in understanding the aforementioned.

{\bf Impact on network engineering:} The data structure underlying the diffusion pattern in a sensor network may exhibit some properties that can greatly impact on the efficiency of the network and traffic engineering processes. In some cases, it can translate into a performance parameter that may be correlated to other key performance parameters of the network. The evaluation of such impact is another key process that can provide important insights on the network engineering process.

The contributions of this paper are threefold. Firstly, we propose a unified and integrated IoT-CPS framework that builds upon a layered model to achieve scalability through modularity. Secondly, we present a diffusion model using an epidemic compartmental 
structure for such a framework. Lastly, a numerical analysis and validation of the efficacy of the proposed diffusion model with respect to network information flow and engineering is presented.

The remainder of this paper is organised as follows. After the current section, we  present the main components of the CPS framework and its underlying inner feedback loop and features in Section \ref{sec:2}. Thereafter, the information diffusion model  is
proposed in Section \ref{sec:3}, while the CPS surveillance model is described  in Section \ref{sec:4}. In Section~\ref{sec:5} the corresponding numerical results are presented while deployment considerations and use cases are described  in 
Section \ref{sec:6}.  Section~\ref{sec:7} draws a conclusion and exposes some avenues for future research based on the results obtained and analysed.  

\section*{The Cyber Physical System Framework}\label{sec:2}
The CPS framework considered in this paper is built around a layered architecture presented in Figure~\ref{cpsf} and a management model that evolves around the inner feedback loop illustrated by Figure~\ref{innerloop}.
\subsection*{The CPS  Layered Architecture}

\begin{figure}[h]
\begin{center}
\includegraphics[scale=0.35]{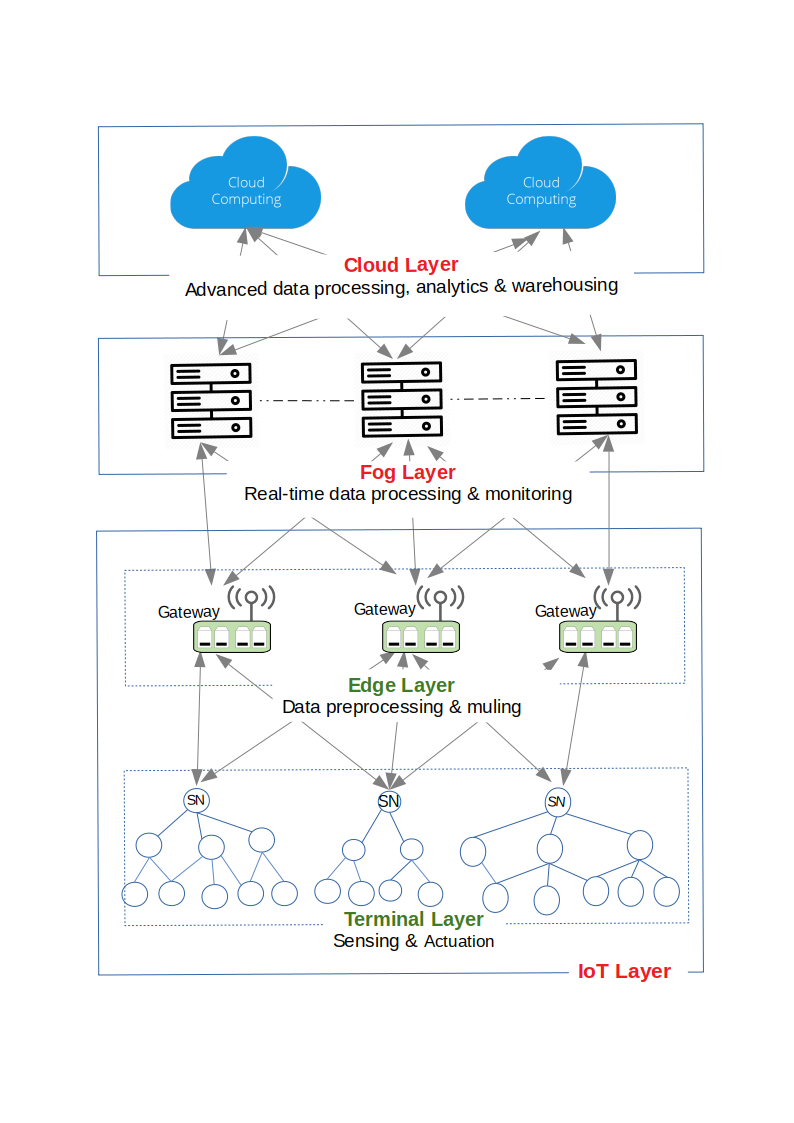} 
\caption{CPS Layered Architecture} \label{cpsf}
\end{center}
\end{figure}

Figure ~\ref{cpsf} depicts our proposed CPS framework. It can be deployed as an IoT-cloud infrastructure that relies on a multi-layer architecture to provide modularity and scalability. These layers are described as follows:

{\bf IoT  Layer:} This layer reveals a platform where sensing devices including positioning, identification and actuation devices are embedded into physical objects to get readings and react on the objects when the obtained readings have reached pre-set thresholds. These sensing devices can be organised as peer-to-peer or tree topologies enabling single hop communication \cite{muling6, muling7} or as a mesh of interconnected devices using a participative approach to complete tasks and multi-hop approach for communication.\cite{1,2,3}. We note that when deployed into a mesh infrastructure, the IoT subsystem will play an important role in the stability, safety and  reliability of the CPS because i) an attack on any node of the subsystem may be masqueraded as a normal reconfiguration process which might have adverse effects on the CPS and ii) the routing process implemented by the IoT subsystem may have a negative impact on the CPS, destabilising it and impacting negatively on its performance. 

{\bf Edge Computing Layer:} The edge computer layer is responsible for analog-to-digital preprocessing and aggregation of data for further processing. While these gateways may be static/fixed, mobile devices such as mobile phones and drones can also be used to collect, preprocess, and ferry/mule the data to processing points as proposed in~\cite{muling1,muling2,muling3,muling4}. This layer also hosts a software defined network (SDN) controller that i) collects routing information from the IoT layer and computes the collection tree used to route the sensor/actuator readings for the IoT subsystem and ii) perform CPS surveillance to discover relevant diffusion patterns in order to apply network reconfiguration in order to re-optimise the CPS under stress conditions.   

{\bf Fog Computing Layer:} The fog computing layer is a storage and processing layer. It is located close to the users to minimise transmission delays, security and ownership and other challenges often associated with public cloud computing infrastructures. 
This layer is responsible for storing and preprocessing the data collected from the edge computing layer before moving the preprocessed data to the cloud for further insights \cite{fog}. 

{\bf Services/Applications Layer:} CPS services are provided to the users through various applications at this layer. They are usually provided using the Software-As-A-Service (SAAS) model of cloud computing. Some of the application areas of the proposed 
CPS model include cyber physical health systems (CPHS)~\cite{health1,health2,health3,health4}, cyber physical solar energy systems (CPSS) ~\cite{energy1,energy2}, cyber physical environment systems (CPES) including water, air 
pollution~\cite{pollution1,pollution2} and noise pollution monitoring systems for the protection of the environment, cyber physical public safety systems (CPPS) ~\cite{safety1,safety2,safety3} and  many other emerging from niche areas such as social 
networks to control how information is disseminated between circles of  friends, relatives, and acquaintances.  

\begin{figure}[h]
\begin{center}
\includegraphics[scale=0.20]{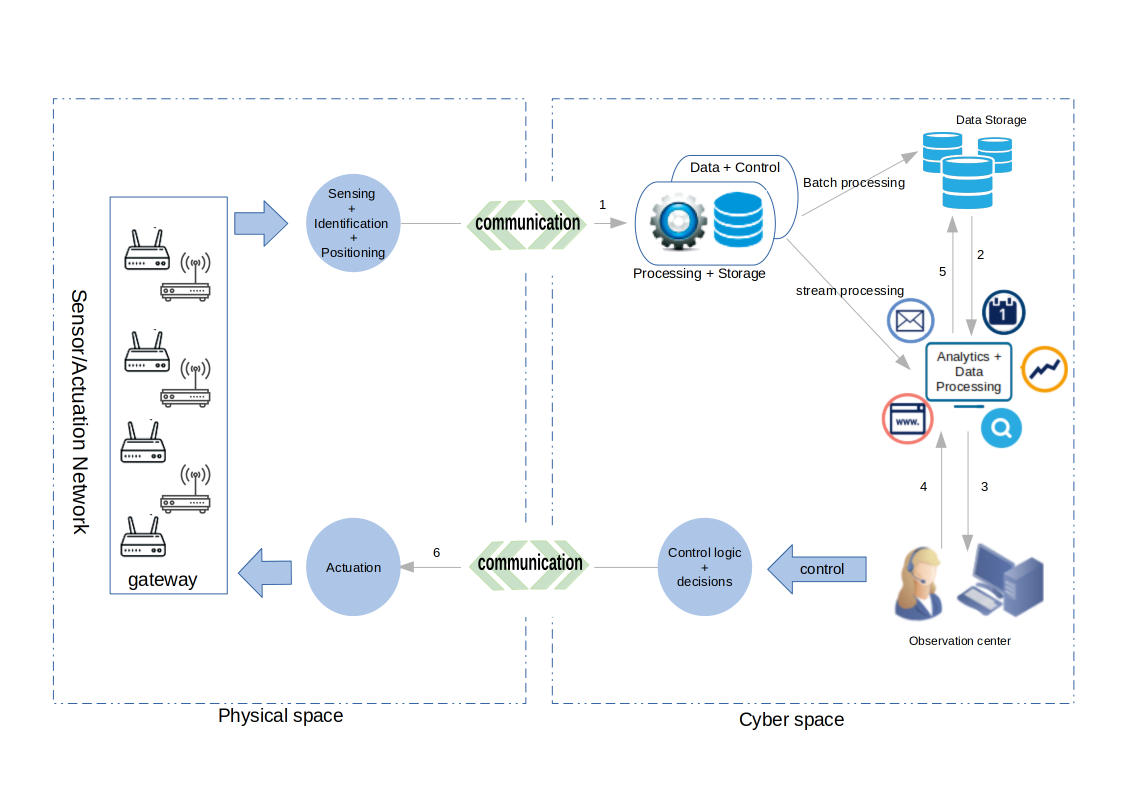} 
\caption{Inner Feedback Loop} \label{innerloop}
\end{center}
\end{figure}

Our proposed framework is made up of four layers but also includes an inner feedback loop revealing the interactions between the cyber and physical spaces. This is shown in Figure~\ref{innerloop} where: 1) a bottom-up flow of information triggered by 
sensing of physical objects in the physical space is translated into services at the higher levels and 2) a top-down flow of information may be initiated by the processing of data in the cyber space to achieve physical actuation on the physical objects managed 
by the CPS.

\subsection*{The CPS Surveillance Model} \label{sec:SAR}

Both the Internet-of-Things and social networks and others are structured around a graph model connecting different nodes. In this graph, edges are the interconnecting links while the vertices are nodes that produce and/or 
consume information. With respect to IoT networks, the nodes can be sensors, actuators, identification nodes (RFID readers and tags) and localisation nodes (GPS nodes); while in social networks, the nodes are usually human users who share the 
information following a structured model embedded into the underlying communication protocols. In the Internet-of-Things, collection tree structures have been massively used as diffusion structures for the collection of information from sensor/actuator nodes 
and disseminating same to sink node(s) for further  processing. On the other hand, social networks (such as Facebook, Whatsapp, Skype, Instagram etc.) might either follow the same diffusion tree structure or a different information dissemination model for sharing 
information between users. 

\begin{figure}[ht]
\begin{center}
\includegraphics[scale=0.30]{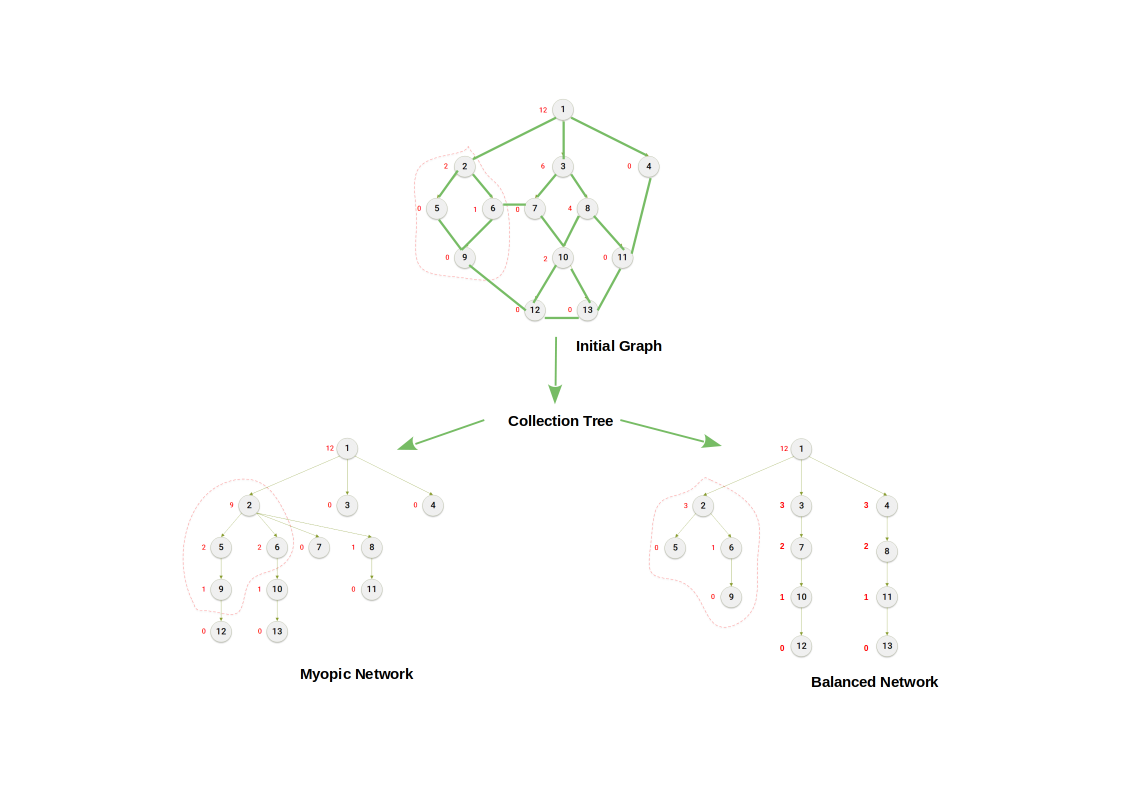} 
\caption{Relevance of CPS Surveillance} \label{graph}
\end{center}
\end{figure}

This paper's focus lies on an information dissemination model that uses collection tree structure for sharing information among nodes of a network. As revealed by Figure~\ref{graph}, an IoT or social network can be structured around a graph model that 
connects nodes (sensors/actuators in an IoT setting or human users in a social network setting) but with different information flowing between them. Information flowing within IoT networks would be sensor/actuator readings while across social networks it 
might be news shared among friends/family members and other users. 

In the IoT, protocols such as RPL~\cite{RPL}, CTP~\cite{CTP} and LIBP~\cite{3b} use collection tree protocols for information collection (from regions of interest) and routing to processing locations (where they are consumed and/or processed). In social 
networks, information dissemination may follow a similar structure where each user is provided a diffusion tree similar to a collection tree to be used for sharing  information among friends/relatives and other social media relationships. Irrespective of the 
application domain, the structure of the diffusion/collection tree might have a great impact on the performance of the underlying network. As illustrated in Figure ~\ref{graph}, two different structures could be derived from the initial network: i) a myopic structure 
that may lead to overloading some nodes (such as node 2 carrying 9 children for example) while leaving others idle or underloaded (node 3 for example carrying only one child/descendant) and ii) a balanced network where each node has relatively lower  
number of dependents/children. 

When applied to an IoT network, the collection tree structure may lead to the IoT provider's network experience engineering performance issues, especially when the node carrying a high number of descendants is a critical node (such as a sensitive unit of an 
hospital for instance controlled by nodes 2, 5, 6 and 9 in our illustration). Similarly, within social networks, when similar information dissemination structures are used, providers might experience greater losses when heavily loaded nodes either fail or are 
attacked. From a user's perspective, these heavily loaded nodes are synonymous to social media influencers, with millions of followers or friends and as such are pivotal in disseminating both good and bad information. This reveals the relevance of i) the impact of the IoT subsystem on the CPS ii) the CPS surveillance to analyse the IoT subsystem in order to take appropriate measures when the CPS reliability and/or its performance is at stake and iii) the CPS reconfiguration for restoring the CPS stability/performance when it is under stress.

The focus of this paper lies on the CPS surveillance using collection tree structures and their potential extensions to social media networks. The IoT subsystem reconfiguration when it is under stress conditions is also an important feature of the CPS, however it is beyond the scope of this paper.

\section*{The Information Diffusion Model} \label{sec:3}
As revealed by Figure~\ref{cpsf}, the CPS of the future will be designed around a network infrastructure that interconnect islands of IoT subsystems  alongside an information diffusion model implemented based on a tree-like topology rooted at gateways. 
The management of such a forest of interconnected trees requires formal modelling and accurate analysis before deployment. The information diffusion model adopted by this paper is presented below.  

\subsection*{\bf Epidemic Models}
Epidemic models known as Susceptible, Infected  and Recovered (or removed) (SIR) were pioneered around 1927 to describe the interaction between individuals when a disease breaks out within a given population~\cite{8}. Since inception, the initial SIR model has been continuously used to solve problems in various domains as recorded by Table~\ref{models}. These include network  problems~\cite{9,10,11,12}. It has also been modified and extended into different other models such as "E-SIRS", "SIRS" 
and "SIR-M". As revealed by Table~\ref{models}, in some of these extended models~\cite{7,9,10}, the network nodes are grouped into disjoint sets leading to compartmental models. The analysis  of these models is usefully done by using the basic 
reproduction  number $R_0$ which is usually computed using the next generation matrix denoted by $K$~\cite{4,5}. These epidemic models can have a tremendous impact on the stability of a network. However, it can be seen from Table~\ref{models} that 
most of these models fall short when considering the impact the death of potentially susceptible, infected and removed individuals has on the network. Moreover, all the models assume a predefined nodal interaction and this can lead to uncontrolled epidemic 
transmission with significantly negative impact on the accuracy of the models.

\begin{table}[h]
\begin{center}
\begin{scriptsize}
\caption{Epidemic Models}
\label{models}
\tiny
\centering
\begin{tabular}{|p{1.5cm}|p{1.6cm}|p{1.5cm}|p{1.6cm}|p{1.5cm}|}
 \hline
 \textbf{Model} & \textbf{Domains} & \textbf{Advantages} & \textbf{Disadvantages} & \textbf{Applications} \\ 
 \hline
Susceptible-Infected-Protected (SIP)
 \cite{1a} & Network security & Evolution of the network is considered as a continuous time Markov Process. &  Nodes are assumed to be similar. & Network Security \& Computer virus detection. \\ 
 \hline
Susceptible-Infected-Susceptible (SIS)
 \cite{2a} & Graph theory \& Mathematical biology & Provide stochastic model using Markov Process. & Nodes in the graph are equal. & Network topology design \& Network phenomena (information dissemination).  \\
 \hline
Susceptible-Infected-Recovery-Susceptible (SIRS)
\cite{3a} & Communication & Simulates real life scenarios.  &  Uncontrolled transition. & Social networking \& Information diffusion. \\
\hline  

Epidemic Routing model 
\cite{4a}&  Routing  &  Support for mobile networks. & All nodes are assumed to be the same. & Mobile \& general network routing. \\ 
\hline 
 
Susceptible-Infected-Recovered with Maintenance (SIR-M)
\cite{5a}&  Epidemiology &  Network flexibility analysis.  & Combined both random and uniform node distribution. & Mobile \& general network communication. \\ 
\hline  
\end{tabular}
\end{scriptsize}
\end{center}
\end{table} 

\begin{figure}[h]
\begin{center}
\includegraphics[scale=0.30]{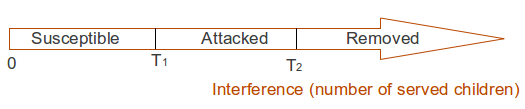}
\caption{Interference thresholds defining epidemic sets} \label{intx}
\end{center}
\end{figure}

When  dealing with collection tree protocols, the number of children carried by a node, also referred to as the node's interference, can be translated into an epidemic state expressing the level of contamination of the node. Using two interference thresholds 
$T_1$ and $T_2$ and following the SIR epidemic model, we consider the SAR model that uses three epidemic states referred to as susceptible (safe), attacked or removed statuses. The safe, attacked and removed states are loosely respectively equivalent 
to the susceptible, infected and removed states of the SIR model. These states and the associated thresholds are depicted by Figure \ref{intx}.

We  define the considered states  as follows:

\begin{enumerate}

\item \textbf{Susceptible nodes:}  are the least or non-interfering nodes in a network. Their total number is denoted by $S$. Each susceptible node $n$  
is assumed to have weight (level of interference) less than the threshold $T_1$. 

\item \textbf{Attacked nodes:} are highly interfering but still operational nodes. The total number of infected nodes in  a network is 
denoted by $A$. An infected node  is assumed to have weight   less than the threshold $T_2 $ but at least equal to the threshold $T_1$. 

\item  \textbf{Removed nodes:} are nodes which are no longer functional because of the high level of interference between them. These nodes are referred to as dead nodes and their total  number is denoted by $R$. A  node  is considered to be removed if its interference   is at least equal to the threshold $T_2 $. 
\end{enumerate}

As defined above, the threshold setting process can be used to partition the network nodes into ``susceptible: $\mathcal{S}$'', ``affected: $\mathcal{A}
$'', and ``removed: $\mathcal{R}$'' epidemic sets which may have different economic and engineering impacts on real networks.

\subsection*{\bf Diffusion Sets}\label{net_def}

Consider a sensor network defined by the structure $G({\mathcal L},{\mathcal N},{\mathcal W}, s)$  where $\mathcal L$ stands for the set of its links, 
$\mathcal N$ is the set of its sensor nodes and $\mathcal W$ is the set of the nodes weights (node interference in LIBP context) while $s$ is the sink of the sensor network. Besides being partitioned into epidemic sets based on interference, the network $G$ can also be partitioned into interference sets which define the way a group of nodes can impact other nodes upon an increase or decrease in interference level, while a collection routing protocol such as LIBP is adopted. In the rest of this paper, we refer to the superset of diffusion sets as $\mathcal{I}=\cup_{\imath} I_{\imath}$. 
\begin{description}\label{def}
\item {\bf Definition:} Consider a network $G({\mathcal L},{\mathcal N},{\mathcal W}, s)$ as defined in Section \ref{net_def} and define $d$ as a distance function where $d(n)$ is the smallest number of hops from the network node $n$ to the sink $s$. A diffusion set $I_{\imath} \in \mathcal{I}$ is a non empty subset of $\mathcal{I}$ satisfying the following properties:
\item {\bf Properties:}
\begin{enumerate}

\item[\textbf{$P_1$:} ]  All nodes in set  $I_{\imath}$ are at the same distance from the sink, i.e. 
$$\forall x, y \in I_{\imath}, \  \ d(x)=d(y)$$

\item[\textbf{$P_2$:} ]  $I_{\imath}$ is a singleton, or for each node $x$ in $I_{\imath}$ there is  another node $y$ in $I_{\imath}$ such that $x$ and $y$  share the next neighbour. (Here,  the next neighbour node of node $n$ refers to a node connected to $n$ which is at $(d(n)+1)^{th}$ hop)

\item[\textbf{$P_3$:} ] For each node $x$ in $N$, if $x$ shares a next neighbour (which is at $d(x)+1$) with some node in $I_{\imath}$, then $x : I_{\imath}$. 
\end{enumerate}
\item {\bf Note:} A diffusion set is determined only by the network structure but might have an impact on the dynamics on the algorithm only upon interference transmission between diffusion sets.
\end{description}

\subsection*{\bf Illustration} 

\begin{figure}[h!]
\centering
\subfigure[A random network]{
\includegraphics[width=0.27\textwidth]{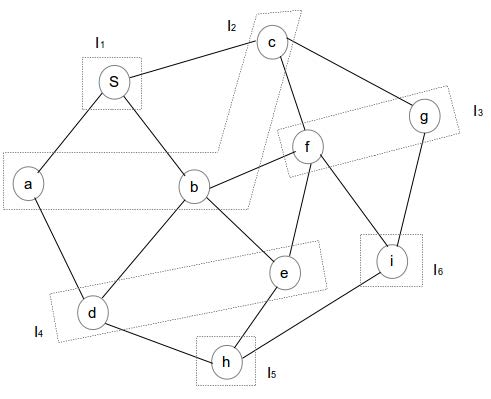}
\label{isetrand}
}
\subfigure[A grid network]{
\includegraphics[width=0.23\textwidth]{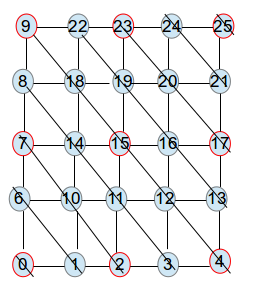}
\label{isetgrid}
}
\caption{Diffusion sets.} \label{intset}
\end{figure}

As an example, Figure \ref{intset} shows two graphs whose nodes are grouped into diffusion sets. 

\subsubsection*{\it\bf Diffusion sets} 

The network depicted by Figure~\ref{isetrand} has a random structure with nodes $a$, $b$ and $c$ being at the same distance (same height) from the sink $s$ and nodes $a$ 
and $b$ sharing the same next neighbour node $d$. Nodes $b$ and $c$ share the same next neighbour node $f$. Therefore $a$, $b$ and $c$ are in the same 
diffusion set (from $P_2$). Likewise, nodes $d$ and $e$ are the same distance from sink $s$ and share the same next neighbour $h$; while nodes $f$ and $g$ are the 
same distance from sink $s$ and share the next neighbour $i$. From Figure~\ref{isetrand}, it can be seen that the nodes are grouped in diffusion sets as follows:

$I_1=\{ S\}$,  $I_2=\{a, b, c\}$,  $I_3=\{f, g\}$, \\
 $I_4=\{d, e\}$,  $I_5=\{h\}$ and $I_6=\{i\}$.
 
It is clear that the set of all nodes $N$ of the presented graph is the union of all diffusion sets of the graph. That is $N=\cup_{i=1}^{6}I_i$. 
On the other hand all diffusion sets  of the graph are pairwise disjoint. That is $I_i \cap I_j\neq \emptyset \Leftrightarrow i = j$. Therefore, the set
${\mathcal I} = \{I_1, I_2, I_3, I_4, I_5, I_6\}$ of all diffusion sets forms \textbf{a partition} of the set $N$. On the other hand, nodes $e$ and $f$
are the same distance from $s$ but do not share the same next neighbour node and are not consequently in the same diffusion set. Furthermore, nodes $h$
and $i$ which do not have a common next neighbour node  are referred to as singletons.

The grid network in Figure~\ref{isetgrid} presents a symmetric structure where all nodes which at the same distance from the sink (Node $0$) are in the
same diffusion set. In that network, the diffusion sets computed are:
  
$I_1=\{0\}, I_2=\{25\}, I_3=\{1, 6\}, I_4=\{2, 10, 7\}, 
I_5=\{3, 11, 14, 8\}, I_6=\{4, 12, 15, 18, 9\}, I_7=\{24, 21\}
I_8=\{22, 19, 16, 13\}, I_9=\{20, 23, 17\}$.

Similarly to the random network, the diffusion sets of the symmetric network form disjoint groups. Furthermore, it is clear that each node belongs to an interference set which partitions the underlining network. This makes it possible to quantitatively model the network system without counting nodes more than once.

It also reveals a compartmental structure where nodes in the same compartment (diffusion set) might experience similar impact from attacks, failures or congestion.  

This shows that the interference transmission on a network would be determined bases on related dynamics across its interference sets.

\subsubsection*{\it\bf Interference Transmission}

\begin{figure}[h!]
\begin{center}
\includegraphics[scale=0.35]{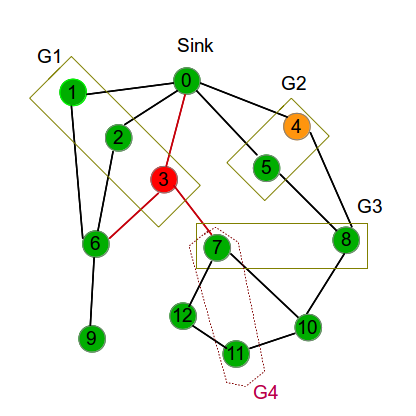} 
\caption{Interference transmission.} \label{s1}
\end{center}
\end{figure}

Given the two networks depicted by Figures~\ref{isetrand} and~\ref{isetgrid}, it can be observed that two nodes are in the same diffusion set if and only if, any increase in interference level of one of them might also result in an increased interference of the other. This means that two nodes can transfer interference to each other. This is possible since these nodes are at the same distance (same height) from the sink and also share the same next neighbour node from the sink. Furthermore, Figure~\ref{s1} illustrates the transmission of interference between diffusion sets by showing that an increase in the interference level of a node does not necessarily affect all nodes in the network but rather affects the set of nodes in its diffusion set.

Figure \ref{s1} shows a network partitioned into diffusion sets $G1$, $G2$, $G3$ and  singletons with node $4$ being attacked while node $3$ is in 
removed state. Note  that  removal may be  caused by an intruder getting access to  network and increasing the node's interference level beyond the $T_2$ 
threshold. In this network, node $5$ which is in the same diffusion set as node $4$ is susceptible to the infection caused by its attacked diffusion set  mate (node $4$). Since node $3$ is removed, all adjacent links should be removed, that is the links $(3,6), (3,7)$. This should lead to node $7$ changing its set membership to form a new diffusion set $G4$ with the singleton $11$. 

Note that the diffusion sets define a new structure of nodes in a network which reveals how an increase of interference level (weight) of a node may impact other nodes. This is why in this case we say that {\it interference is transferred from one node to another}.

\section*{The CPS Surveillance Model}\label{sec:4}

As described above, we consider a diffusion model where nodes are grouped into diffusion sets with nodes in the same diffusion set assumed to be infectiously similar to each others while those in different diffusion sets behave differently. This section presents an analytical model of the interference diffusion and an analysis of its stability. 

In our model, each diffusion set $I_{\imath}$ is composed by subsets of nodes defined by $I_{\imath} =\{S_{\imath}, A_{\imath}, R_{\imath}\}$. The subset $S_\imath$ of {\bf susceptible nodes} which are working normally and not yet affected is of a size denoted by $S_i$ while the subset {\bf attacked nodes} in $I_{\imath}$ has its size defined by  $A_i$. Finally the subset of {\bf removed nodes} in $I_{\imath}$ is of size defined by $R_i$.  When the diffusion set $I_\imath$  includes both the attacked and removed sets which are non empty ($A_i \neq 0$  or $R_i \neq 0$), it is said to be an \textbf{infected set}. The diffusion problem consists of finding for each diffusion set $I_\imath$ its evolution function over time $I_\imath (t) =\{S_i(t), A_i(t), R_i(t)\}$ as defined below
\begin{equation} \label{eq:diff}
\begin{array}{ll}
Find & I_\imath (t)= [S_i(t), A_i(t), R_i(t)]  \\
subject & \mbox{ to } 
\end{array}
\end{equation}

$$
\left \{
\begin{array} {llr}
\forall x \in {\mathcal S}  & \rightarrow w(x) < T_1 &\mbox{     (1.a)}\\
\forall x \in {\mathcal A} & \rightarrow T_1 \leq w(x) < T_2 &\mbox{     (1.b)}\\  
\forall x \in {\mathcal R} & \rightarrow w(x) \ge T_2 &\mbox{     (1.c)} \\ 
\forall \mathcal{S}_\imath,\mathcal{A}_\jmath  \in {\mathcal I} & \rightarrow \lambda_{ij}  \le \lambda &\mbox{     (1.d)} \\
\forall \mathcal{A}_\imath,\mathcal{R}_\jmath  \in {\mathcal I} & \rightarrow \rho_{ij}  \le \rho &\mbox{     (1.e)} \\
\forall \mathcal{S}_\imath  \in {\mathcal I} & \rightarrow \eta_i  \le \eta &\mbox{     (1.f)} 
\end{array}\right.
$$

where  $\lambda_{ij}$ is the transmission rate from susceptible diffusion set $\mathcal{S}_\imath$ to attacked diffusion set $\mathcal{A}_\jmath$ while $ \rho_{ij}$ is the transmission rate from attacked diffusion set $\mathcal{A}_\imath$ to removed diffusion set $\mathcal{R}_\jmath$. $\eta_i$
is a parameter revealing the impact of the diffusion on the network if a susceptible node in diffusion set  $\mathcal{S}_\imath$  becomes infected: attacked or removed. Note that the diffusion model does not express any  dependability constraints. It only expresses a set finding function and how it is 
mapped into i) {\it a set finding problem} expressed by the routing objective~(1). Equations~(1.a),~(1.b) and~(1.c) express the network partitioning into interference states: susceptible: $\mathcal S$, attacked:  $\mathcal A$ and removed: $\mathcal R$. Equations~(1.d),~(1.e) and~(1.f) are diffusion equations that express and limit the interference diffusion from one set to another and reveal the impact of moving nodes from the susceptible to infected states (attacked or removed) on the network. 

\subsection*{\bf Analytical Description}

\begin{figure}[h!]
\begin{center}
\includegraphics[scale=0.45]{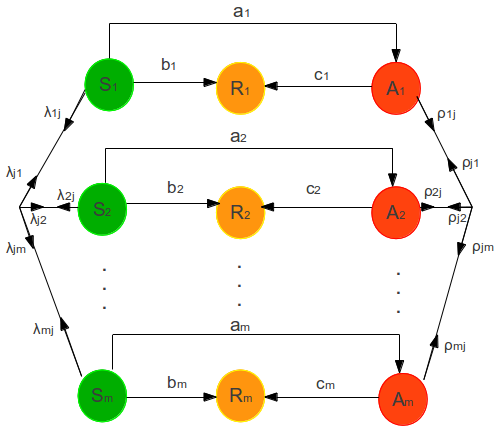}
\caption{Representation of interference transmission.} \label{ss2}
\end{center}
\end{figure}

Figure~\ref{ss2} presents a finite state machine of our diffusion model for a network which has been partitioned into  $m$ diffusion sets. It reveals the states of the diffusion sets and for each state its associated transitions as a well as the actions that trigger transitions from one state to another. 

Susceptible nodes in the diffusion set $I_\imath$  may be attacked at the rate $a_i$, while the attacked nodes from $I_\imath$ get removed at the rate $c_i$. Susceptible nodes in the interference  set $I_\imath$ may  highly increase their interference levels so as to move directly to the removed status without transiting via the attacked status. On the other hand removed nodes my cause some of the susceptible or attacked nodes to leave the network
because of the destruction of links.  We consider  $bi$ to be the rate with which, susceptible nodes in $I_\imath$ are removed. Susceptible nodes from diffusion set $I_\imath$  migrate to diffusion set $I_\jmath$  with rate $\lambda_{ij}$, and infected nodes in $I_\imath$  migrate to $I_\jmath$ at a rate of $\rho_{ij}$. Note that in our model, removed nodes may cause migration of nodes from one diffusion set to another. The difference equation (equation~\ref{mod}) presented below is a result of the assumptions made above. 

\begin{equation}\label{mod}
 \begin{cases}
  S_i'= -a_i S_i+ \sum\limits_{j\neq i}\lambda_{ji}S_j- \sum\limits_{j\neq i}\lambda_{ij}S_i-b_iS_i  \\
  
  A_i'= a_i S_i+ \sum\limits_{j\neq i}\rho_{ji}A_j- \sum\limits_{j\neq i}\rho_{ij}A_i-c_iA_i  \\
   
  R_i' = b_iS_i+c_iA_i \\
  \end{cases}
\end{equation}
Note that $S_i$, $A_i$ and  $R_i$  are functions of time $t$ for each diffusion set $I_\imath$. The negative rates in the model represents a decrease, while the positive ones represent an increase. The parameter $a_i$ stands for the transmission rate between susceptible  and infected nodes. This parameter depends directly on the number of susceptible nodes $S_i$  and the infected ones $A_i$. It therefore makes sense to relate $a_i$ with two other measures:
\begin{enumerate}
\item The susceptibility rate of each node in the diffusion set $I_\imath$, denoted by $\beta_i$.
\item Infectiousness rate of  nodes in the infected diffusion set $I_\imath$, denoted by  $\gamma_i$.
\end{enumerate}
Conversely, the structure of a diffusion set clearly influences the attack ability, as the effects of infections vary across different diffusion sets. We use the parameter  $\eta_i$ as a measure of the impact on the network structure when a node gets infected (attacked or removed).
Therefore, $a_i$ can be computed using the following formula:  
\begin{equation}\label{a}
a_i= \beta_i \gamma_i \eta_i \frac{A_i}{N}
\end{equation}
where $\frac{A_i}{N}$ denotes the fraction of infected nodes in diffusion set $I_\imath$.

By replacing the equation~\ref{a} into the difference equation~\ref{mod}, we get the following equations.

\begin{equation}
 \begin{cases}
 S_i'= -\beta_i \gamma_i \eta_i \frac{A_i}{N} S_i+ \sum\limits_{j\neq i}\lambda_{ji}S_j- \sum\limits_{j\neq i}\lambda_{ij}S_i-b_iS_i  \\
  
  A_i'= \beta_i \gamma_i \eta_i \frac{A_i}{N} S_i+ \sum\limits_{j\neq i}\rho_{ji}A_j- \sum\limits_{j\neq i}\rho_{ij}A_i-c_iA_i  \\
   
   R_i'  = b_iS_i+c_iA_i \\
  \end{cases}
\label{Model}
\end{equation}

\subsection*{\bf Stability Analysis}\label{as}

In this section, we study  the stability of the  system at the disease-free equilibrium points by assuming that:  
\begin{enumerate}

\item there is no node newly joining the network. That is,  at any time $t$, if $N$ is the number of nodes at time $t=0$, then  $  N= \sum \limits_i S_i(t) \ +\ \sum\limits_iA_i(t)\ +\ \sum\limits_iR_i(t)$.
\item the death ( not caused by interference) and birth rate are assumed to be zero.
\item the rates  in equation~\ref{Model} are constant.
\item in the networks considered, the death of nodes do not cause new diffusion sets formation.
\end{enumerate}

We first compute the disease-free equilibrium of the system which will be used to compute the basic reproduction number $R_0$ which is the number used for studying the stability.

\subsubsection*{\it\bf Disease-free equilibrium (DFE)}

Consider  equation \ref{Model} and assume the sets of diffusion sets in $\mathcal{I}$ whose size is $m$. Since the system is not affected by the number of removed nodes $R$, the equation of $R$ is omitted. We present DFE as 
 $E=(e_1, e_2, \cdots, e_{2m})=(S_i, A_i=0)\ , i=1,2,..., m$,   which verifies the equations
 \begin{equation} \label{eq1}
  \forall i\in {\mathcal I}, \ 
 \beta_i \gamma_i \eta_i \frac{A_i}{N} S_i+ \sum\limits_{j\neq i}\lambda_{ji}S_j- \sum\limits_{j\neq i}\lambda_{ij}S_i-b_iS_i =0.
 \end{equation}
Since $A_i=0$, Equation~\ref{eq1} is reduced to:

\begin{equation}\label{eq}
\forall i \in {\mathcal I}, \ 
 \sum\limits_{j\neq i}\lambda_{ji}S_j - \sum\limits_{j\neq i}\lambda_{ij}S_i-b_i S_i = 0.
\end{equation}

Since the system of equations~(\ref{eq}) is linear, it can be written in matrix form

\begin{equation}\label{mat}
SA=0
\end{equation}
where, $ S=(S_1  S_2 \cdots S_m )$ and 
$$A =
 \begin{pmatrix}
  -\sum\limits_{j\neq 1}\lambda_{1j}-b_1 & \lambda_{12} & \cdots & \lambda_{1m} \\
  \lambda_{21} & -\sum\limits_{j\neq 2}\lambda_{2j}-b_2 & \cdots & \lambda_{2m}  \\
  \vdots  & \vdots  & \ddots & \vdots  \\
  \lambda_{m1} & \lambda_{m2} & \cdots & -\sum\limits_{j\neq m}\lambda_{mj}-b_m
 \end{pmatrix}
$$

\begin{itemize}
\item[\textbf{Case 1:}]If  $detA\neq 0 $ then $S_i=0$, $i=1,2,\cdots m$ is the unique solution of Equation (\ref{mat}).
In this case the DFE is $E_0=(S^*_i=0, A^*_i=0)$.
\item[\textbf{Case 2:}] If  $detA = 0 $ then the system of equations ( \ref{mat}) has infinitely many solutions, and thus the system will have infinite number of DFE whose form is $E=(S_i^*, A_i^*=0)$ where $S_i^*$ may not all be zero. 
\end{itemize}
Note that according to Linear Algebra, $det(A)=0$ if and only if the rows or columns of $A$ are linearly \textbf{dependent}.
This can help us to study the dependency of diffusion sets in terms of interference transmission.

\subsubsection*{\it\bf Stability of a network at DFE}

We can study stability using the basic reproduction number $R_0$.
$R_0$ can be calculated using the next-generation matrix approach as described in  \cite{4,5}.

After removing the $R_i$ equations, the systems of equation~(\ref{eq}) left can be decomposed into two subsystems as follows:

\begin{equation}
\mathcal{F}_i(S_i, A_i)=
 \begin{cases}
   0 \\
  
   \beta_i \gamma_i \eta_i \frac{A_i}{N} S_i \\
  \end{cases}
\label{f}
\end{equation}

\begin{equation}
\mathcal{V}_i(S_i, A_i)=
 \begin{cases}
 \beta_i \gamma_i \eta_i \frac{A_i}{N} S_i- \sum\limits_{j\neq i}\lambda_{ji}S_j+ \sum\limits_{j\neq i}\lambda_{ij}S_i+b_iS_i  \\
  
 -\sum\limits_{j\neq i}\rho_{ji}A_j+ \sum\limits_{j\neq i}\rho_{ij}A_i+c_iA_i \\
  \end{cases}
\label{model}
\end{equation}

The next-generation matrix is $K=FV^{-1}$ where $F$ and $V$ are the Jacobian matrices of $\mathcal{F}$ and $\mathcal{V}$ respectively, evaluated at the DFE.
\begin{itemize}
\item[\textbf{Case1:}]If  $detA\neq 0 $  the DFE is $E_0=(S_i^*=0, A^*_i=0)$.
and the Jacobian of $\mathcal{F}$  evaluated at $E_0$ is 

$F_{ij}(E_0)=\frac{\partial\mathcal{F}_i}{\partial e_j}(E_0)$ is the zero matrix.

Consequently, the matrix  $K=FV^{-1}$ is the zero matrix. The eigenvalues of  the matrix $K$ are all zero and hence the basic reproductive number is $R_0=0$. 
Since $R_0<1$, the DFE $E_0$ is globally stable. This is explained by the fact that at $E_0$ the network  is empty and will remain empty because no new nodes join it.

\item[\textbf{Case2:}] If  $detA = 0 $ then the system of equations (\ref{mat}) has more than one  solutions, and thus the system will have more than one  DFE  points whose form is $E=(S_i^*, A_i^*=0)$ where $S_i^*$ may not all be zero. 

\begin{center}
$F =
 \begin{pmatrix}
  \beta_1 \gamma_1 \eta_1 \frac{S_1^*}{N}  & 0 & \cdots & 0 \\
  0 & \beta_2 \gamma_2 \eta_2 \frac{S_2^*}{N}& \cdots & 0  \\
  \vdots  & \vdots  & \ddots & \vdots  \\
  0 & 0 & \cdots & \beta_m \gamma_m \eta_m \frac{S_m^*}{N}
 \end{pmatrix}=  [\delta_{ij}(\beta_i \gamma_i \eta_i \frac{S_j^*}{N} )]_{ij}$

 $V =
 \begin{pmatrix}
\sum\limits_{j\neq 1}\rho_{1j}+c_1& 0 & \cdots & 0 \\
  0 & \sum\limits_{j\neq 2}\rho_{2j}+c_2& \cdots & 0  \\
  \vdots  & \vdots  & \ddots & \vdots  \\
  0 & 0 & \cdots & \sum\limits_{j\neq m}\rho_{mj}+c_m
 \end{pmatrix}=  [\delta_{ij}(\sum\limits_{j\neq 1}\rho_{1j}+c_i)]_{ij}$
 \end{center}
where $\delta_{ij}$ is the Kronecker delta. That is
$\delta_{ij}=(i=j)$.

$$K=FV^{-1}=[(\frac{\beta_i \gamma_i \eta_i \frac{S_j^*}{N} }{\sum\limits_{j\neq i}\rho_{ij}+c_i})\delta_{ij}]_{ij}$$
Since $K$ is a diagonal matrix, the basic reproduction number is $$R_0=Trace(K)=\sum \limits_{i=1}^m \frac{\beta_i \gamma_i \eta_i \frac{S_i^*}{N} }{\sum\limits_{j\neq i}\rho_{ij}+c_i}.$$
\end{itemize}

\section*{Performance Evaluation} \label{sec:5}
This section reports on the numerical analysis of the diffusion model in relation to the diffusion sets and the epidemic sets. Table~\ref{num} shows the initial conditions of the considered network and its performance benchmark parameters.
 
\begin{table}[h]
\begin{scriptsize}
\caption{Diffusion model's parameters.}
\label{num}
\begin{tabular}{ p{2cm} p{5cm} p{5cm} }
\hline
\textbf{Parameter} & \textbf{Description} & \textbf{Value}  \\\hline     
   N   &  Total number of nodes   of a network & 193  \\ \hline
m   &  Number of  chosen diffusion sets & 4  \\ \hline
$S_i^0$  &  Initial number of susceptible nodes in the set i  &  $S_1^0=30$,  $ S_2^0 =30$ , $S_3^0=33$,  $ S_4^0 =35$  \\\hline
$A_i^0$  &  Initial number of infected nodes in the set i   &  $A_1^0=10$,  $ A_2^0 =20$  ,  $A_3^0=25$,  $ A_4^0 =25$    \\\hline
$R_i^0$  &  Initial number of removed nodes in the set i   &   $R_1^0=0$,  $ R_2^0 =0$ , $R_3^0=5$,  $ R_4^0 =0$  \\ \hline
$ \lambda_{ij}$ & Transmission rate from susceptible diffusion set $i$ to infected diffusion set $j$    &  $ \lambda_{12}=0.04$, $ \lambda_{21}=0.03$, $\lambda_{ij}=0$,\\ & & with $i>2$ or $j>2$ \\ \hline
$ \rho_{ij}$ & Transmission rate from infected diffusion set $i$ to removed diffusion set $j$    &  $ \rho_{12}= 0.01=\rho_{21}$, $\rho_{ij}=0$, with $i>2$ or $j>2$\\ \hline
$b_i$  & Migration rate from  susceptible nodes in diffusion set $i$ to infected nodes in $i$  &   $b_1=0.01$,  $b_2=0.02$, $b_3=0.03$,  $b_4=0.04$ \\\hline  
$c_i$  & Migration rate from  infected nodes in diffusion set $i$ to removed nodes in $i$  &  $c_1=0.02$,  $c_2=0.02$, $c_3=0.03$,  $c_4=0.04$  \\\hline
$\beta_i$    &  Susceptibility of a node in diffusion set $i$  & $\beta_1=0.11$, $\beta_2=0.1$, $\beta_3=0.2$, $\beta_4=0.3$  \\\hline
$\gamma_i$  &Infectiousness  of a node in diffusion set $i$  & $\gamma_1=0.4$,   $\gamma_2=0.4$, $\gamma_3=0.5$,   $\gamma_4=0.6$  \\\hline
 $\eta_i$    &  Network impact if a susceptible node in diffusion set  $i$  becomes infected: attacked or removed & $\eta_1=0.4$,   $\eta_2=0.4$, $\eta_3=0.5$,   $\eta_4=0.6$  \\\hline
\end{tabular}
\end{scriptsize}
\end{table} 
The results presented in this section are related to the diffusion model expressed by the difference equation~\ref{mod}. To solve this equation, we used the Euler method described in~\cite{6}. For all our simulation experiments, the numerical results are 
presented as graphs that show the changes in the number of nodes in each class of a given diffusion set. Our numerical computation and plotting of related graphs were carried out using the PyLAB Python package. 
  
\subsection*{\bf The Performance Parameters} \label{idea}

Three main performance parameters are considered: 1) the performance patterns 2) the migration rates from safe state to other states and 3) the migration rates across diffusion sets.

\begin{enumerate}
\item {\bf Performance patterns.} The performance patterns $[S_i(t), A_i(t), R_i(t)]$ are functions that shows the evolution of routing process over time. They reveal how nodes move from states to states and across diffusion sets.  
\item {\bf State-to-state migration.} This performance parameter expresses the rate by which nodes change states by moving from one state to another in the order defined earlier: S, A, R. It includes two parameters: the safe to attacked migration expressed by $s-2-a$ and the attacked to removed migration expressed by $a-2-r$ parameter.  
\item {\bf Set-to-set migration.} This performance parameter expresses the rate of which nodes change their affiliation by moving from one diffusion set to another. It is expressed by $s(x)-2-s(y)$ where $x$ and $y$ are two different diffusion sets. For the 4 states considered, this might lead to 12 combinations. 
\end{enumerate} 

We also consider four diffusion sets namely i)  {\bf  $I_1=(S_1,A_1,R_1)$} ii)  {\bf  $I_2=(S_2,A_2,R_2)$} iii)  {\bf  $I_3=(S_3,A_3,R_3)$} and iv)  {\bf  $I_4=(S_4,A_4,R_4)$}.

\subsection*{\bf Performance benchmark results}

\begin{figure}[ht]
\centering
\subfigure[Susceptible nodes.]{
\includegraphics[scale=0.20]{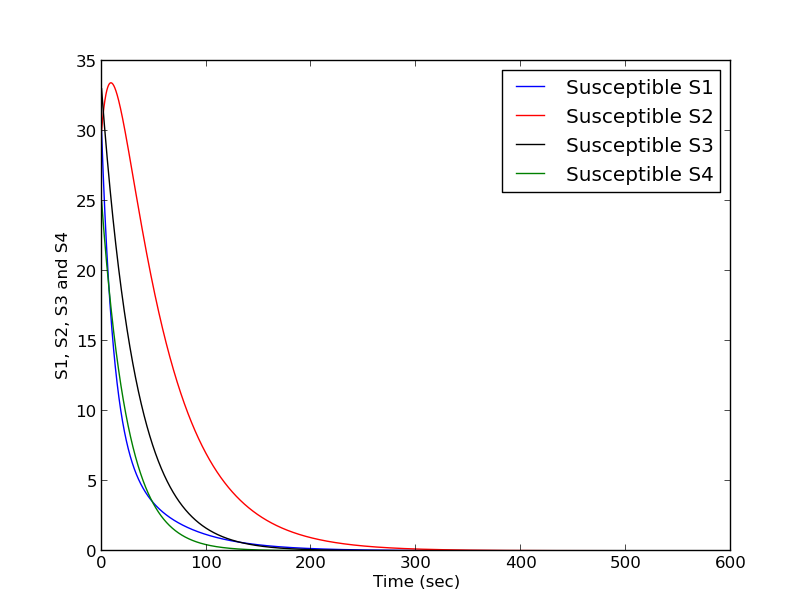} 
\label{s1s2s3s4}
}
\subfigure[Attacked nodes.]{
\includegraphics[scale=0.20]{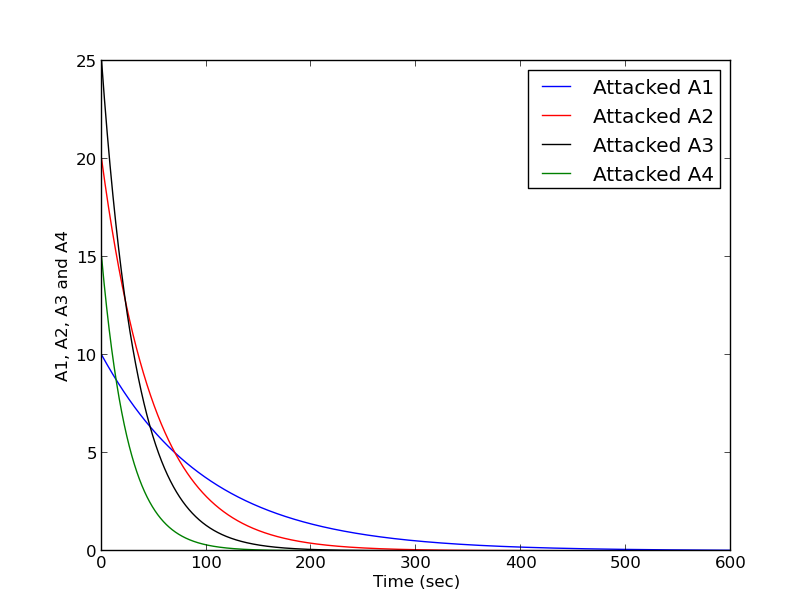} 
\label{a1a2a3a4}
}
\subfigure[Removed nodes.]{
\includegraphics[scale=0.20]{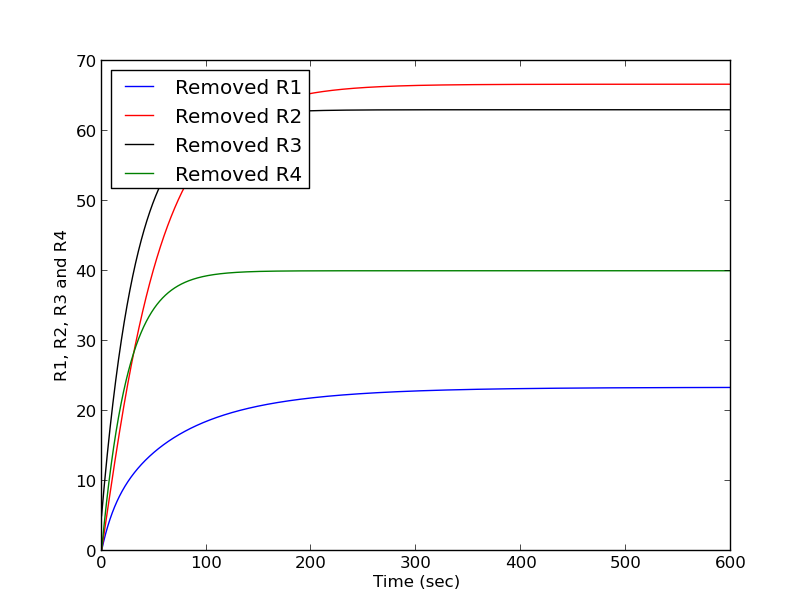} 
\label{r1r2r3r4}
}
\caption{States comparison.}
\end{figure}

{\bf Susceptible states.} Figure \ref{s1s2s3s4} reveals a ``birth-growth-decrease'' pattern where the number of susceptible nodes born when joining the network grows with increase in interference (when the nodes accept children) and decreases when they reach a high interference threshold defining an infected status: attacked or removed. According to the figure, the number of susceptible nodes $S_2$ in diffusion set $I_2$  first increases, then decreases towards zero after a few seconds. In all other diffusion sets the number of susceptible nodes decreases immediately towards zero. The number of susceptible nodes in diffusion set $I_4$ decreases faster than in all other diffusion sets. 

{\bf Attacked states.} Figure \ref{a1a2a3a4} reveals a ``birth-decrease'' pattern where the number of attacked nodes born with the increase of interference in the susceptible sets decrease with time and interference to reach a threshold defining their removal. This is confirmed in the figure by the number of attacked nodes in all diffusion sets which is a decreasing function tending towards zero. The number of infected nodes in diffusion set $I_4$ decreases fastest while those in diffusion set $S_1$ decrease much slower than the others.

{\bf Removed states.} Figure \ref{r1r2r3r4} reveals that the general trend of the removed nodes is of a ``growth-plateau'' where the number of removed nodes initially grow and end up plateauing after a period of time. The figure shows that the number of removed nodes in the four diffusion sets increases until it tends to a non zero value. Note that the removed nodes in the diffusion set $I_1$ does not tend to the total number of nodes in the initial set because some of them might have migrated to diffusion set $I_2$ (see Table \ref{num}). 

\subsection*{\bf Parameters adjustment using diffusion rates}\label{numr2}

In this section, we study the effect of changing the migration rates  by considering the same network (see Table \ref{num}) but with a change in the migration rates of the second diffusion set. The expectation is to evaluate the impact of the adjustment on i) the 
economic efficiency when there is a need to keep the nodes in the attacked state longer in the CPS before reaching the removed state for non-sensitive applications and ii) the engineering efficiency where the focus is on keeping the nodes as long as possible 
in the susceptible state for sensitive applications requiring high QoS.

\subsubsection*{\it\bf Targeting the economic efficiency}

\begin{figure}[hbp!]
\centering
\subfigure[Susceptible nodes.]{
\includegraphics[scale=0.20]{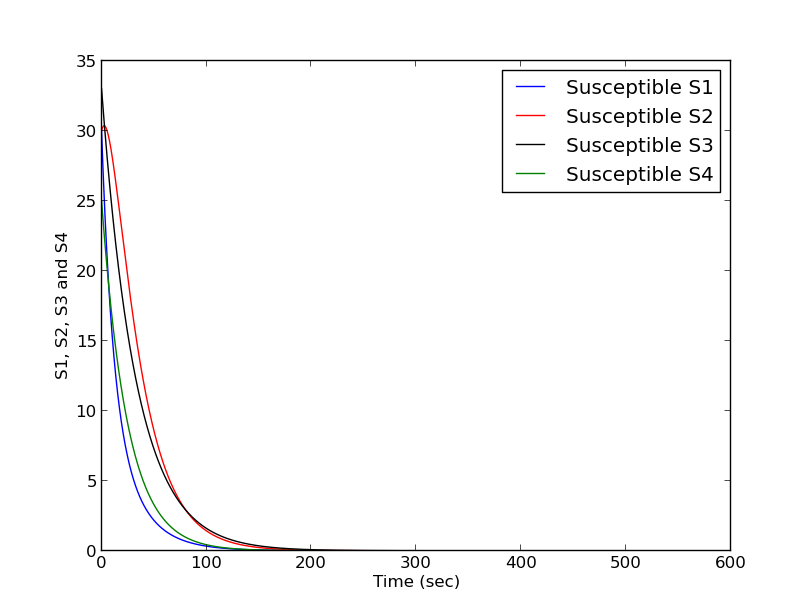} 
\label{o1}
}
\subfigure[Attacked nodes.]{
\includegraphics[scale=0.20]{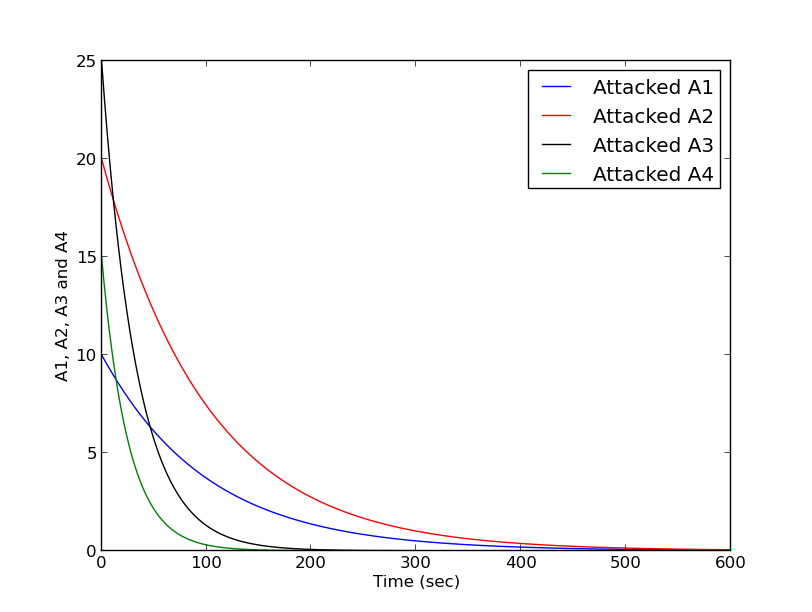} 
\label{o2}
}
\subfigure[Removed nodes.]{
\includegraphics[scale=0.20]{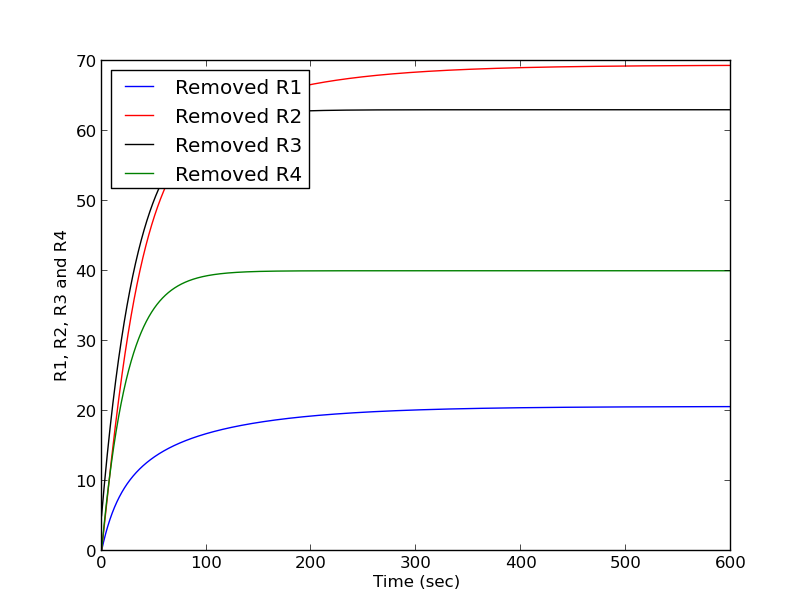} 
\label{o3}
}
\caption{Impact of migration rates on economic efficiency.}
\end{figure}

The rates were changed by lowering the migration rate from susceptible to attacked states by dividing it by 2 (i.e. $b_2'=b2/2=0.01$), while increasing the migration rate from attacked to removed by multiplying it by 2 (i.e. $c_2'=c2\times 2=0.04$). This illustrates a network that requires high economic efficiency, that is the nodes are configured to stay for a shorter time in the susceptible state but stay longer in the attacked state before being removed. Graphs showing the change in the number of nodes in each state are depicted in Figures \ref{o1}-\ref{o3}.

Figure \ref{o1}  shows that due to the change in migration rate, the convergence time (time to 0) for the susceptible nodes $S_2$ in diffusion set $I_2$ is reduced compared with the benchmark case shown by Figure \ref{s1s2s3s4}. In contrast, Figure \ref{o2} reveals an increase of the convergence time of the number of attacked nodes $A_2$ in the diffusion set $I_2$ where the convergence is delayed compared to the case shown by  Figure  \ref{a1a2a3a4}. A similar performance pattern (not reported in this paper for space saving) was observed when similar changes to the migration rates were replicated in the other diffusion sets $I_1$, $I_3$ and $I_4$. Figure \ref{o3} shows that the change in migration rates did not affect the removed nodes neither in the second diffusion set $I_2$ nor in the other diffusion sets when compared to the benchmark scenario depicted by Figure \ref{r1r2r3r4}.

\subsubsection*{\it\bf Targeting the engineering efficiency}

\begin{figure}[h]
\subfigure[Susceptible nodes.]{
\includegraphics[scale=0.19]{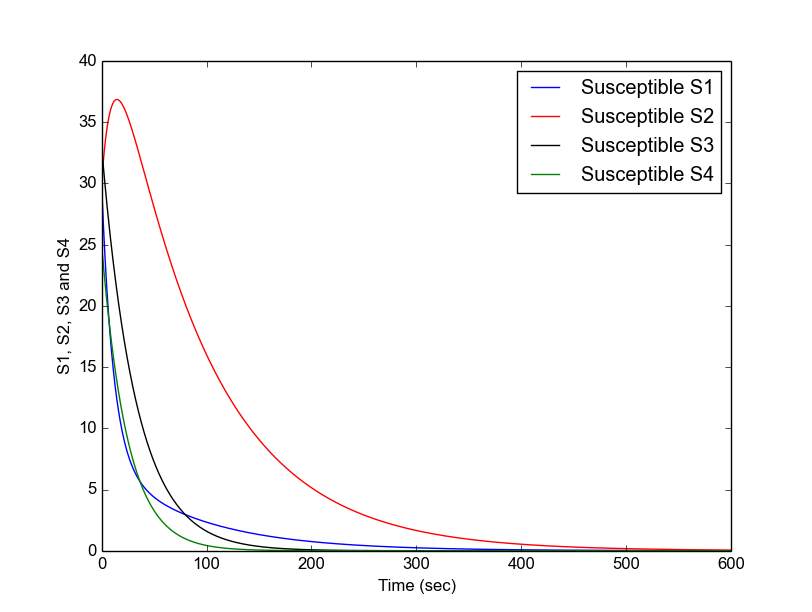} 
\label{oo1}
}
\subfigure[Attacked nodes.]{
\includegraphics[scale=0.19]{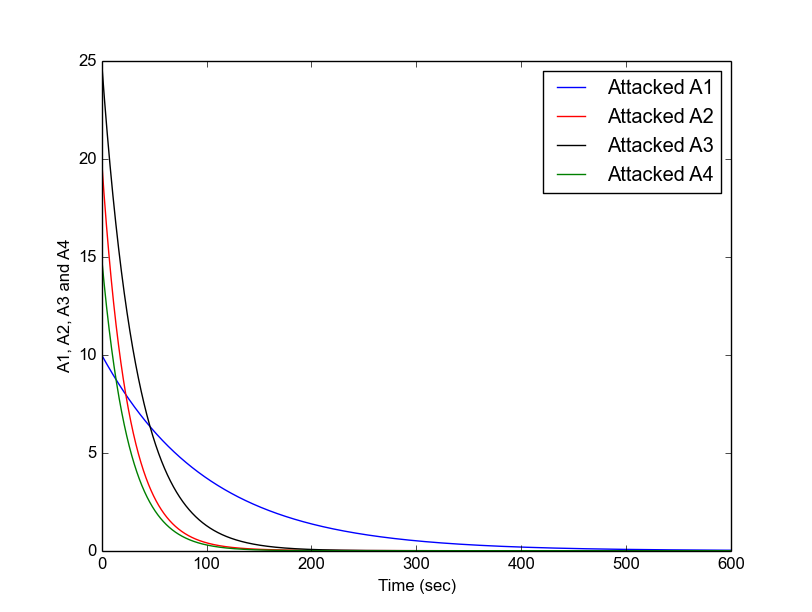} 
\label{oo2}
}
\subfigure[Removed nodes.]{
\includegraphics[scale=0.19]{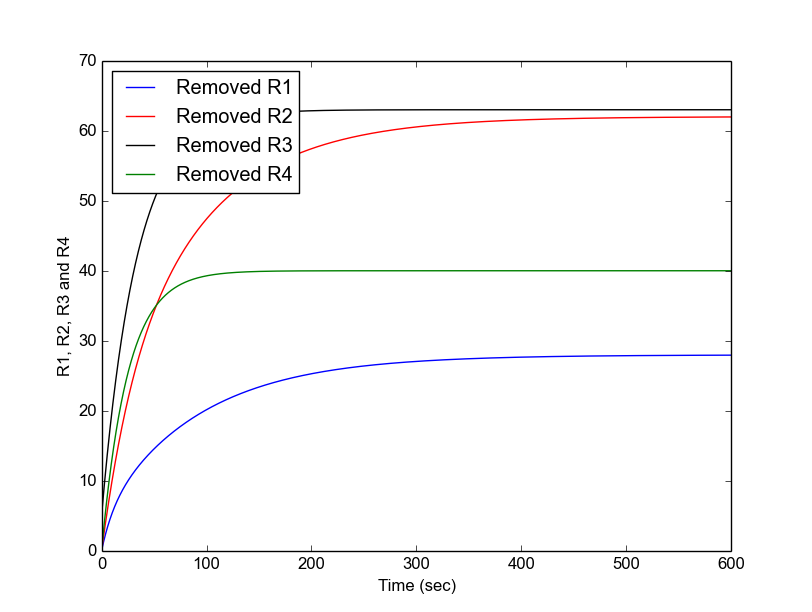} 
\label{oo3}
}
\caption{Impact of migration rates on engineering efficiency.}
\end{figure}

We conducted another set of experiments where the rates were changed by increasing the migration rate from susceptible to attacked states by multiplying it by 2 (i.e. $b_2'=b2\times2=0.04$) while decreasing the migration rate from attacked to removed by dividing it by 2 (i.e. $c_2'=c2/2=0.01$). This illustrates a network that requires high engineering efficiency, that is the nodes are configured to stay longer in the susceptible state but when they move into the attacked state, they quickly move into the removed state. The graphs showing the change in number of nodes in each state are depicted in Figures \ref{oo1} - \ref{oo3} 

Figure \ref{oo1}  shows that due to the change in migration rate, the convergence time (time to 0) for the susceptible nodes $S_2$ in diffusion set $I_2$ is increased compared with the benchmark case Figure \ref{s1s2s3s4}. In contrast, Figure \ref{oo2} reveals a reduction of the convergence time of the number of attacked nodes $A_2$ in diffusion set $I_2$ where the convergence is accelerated compared to the case shown by  Figure \ref{a1a2a3a4}. A similar performance pattern (not reported in this paper for space saving) was observed when replicating similar changes in the other diffusion sets $I_1$, $I_3$ and $I_4$. Like in the economic efficiency in the previous subsection, Figure \ref{oo3} also shows that the change in migration rates did not affect the removed nodes $R_i$ neither in the second diffusion set $I_2$ nor in the other diffusion sets when compared to the benchmark scenario depicted by Figure \ref{r1r2r3r4}.

\subsection*{\bf Parameters adjustment using the network impact}\label{numr4}

\begin{figure}[h]
\centering
\subfigure[Susceptible nodes.]{
\includegraphics[scale=0.20]{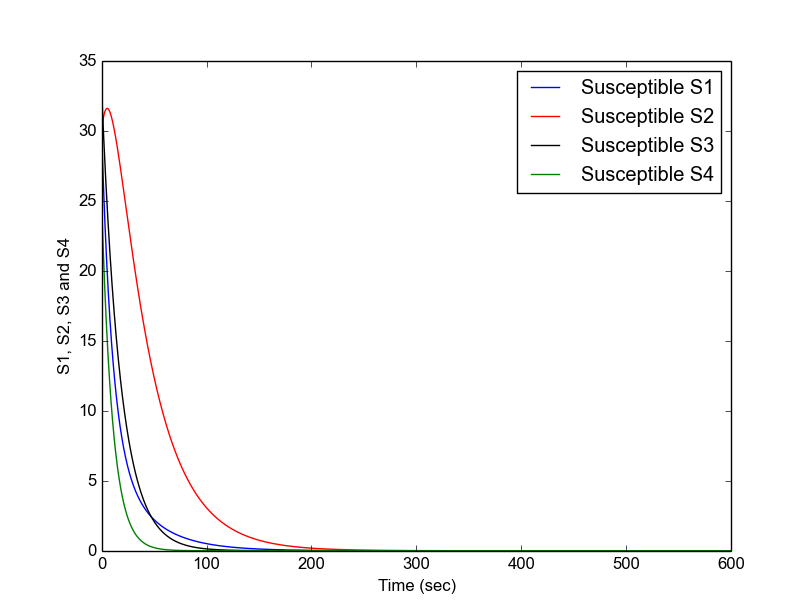} 
\label{p1}
}
\subfigure[Attacked nodes.]{
\includegraphics[scale=0.20]{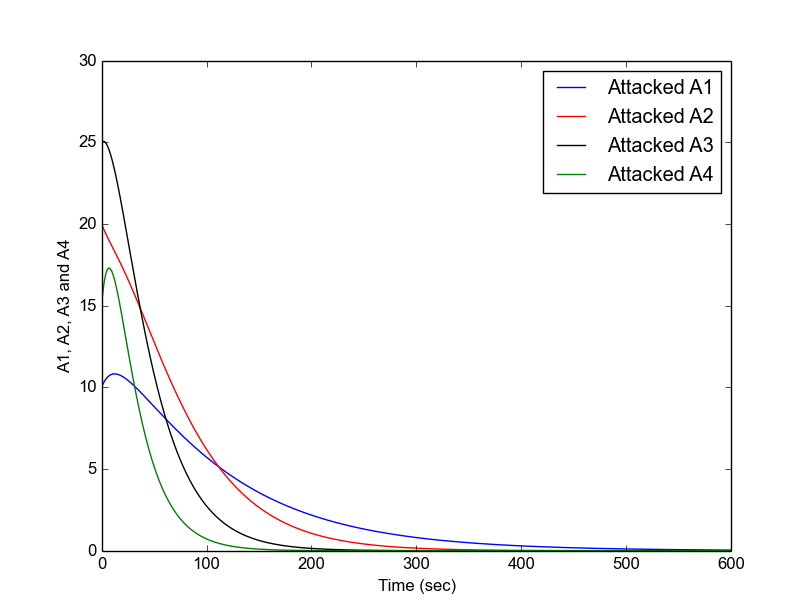} 
\label{p2}
}
\subfigure[Removed nodes.]{
\includegraphics[scale=0.20]{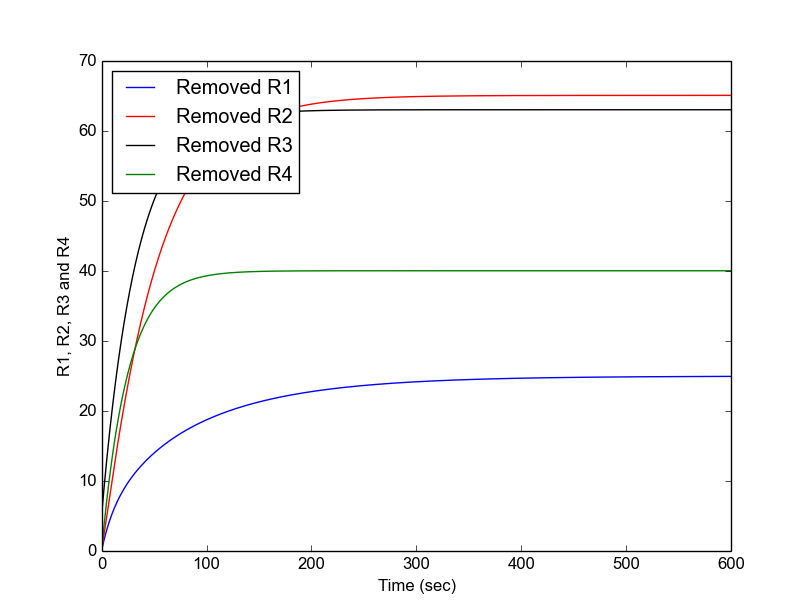} 
\label{p3}
}
\caption{Network Impact Factor: $100*\eta_2$.} \label{netimp}
\end{figure}

We increased the  network impact parameter by multiplying it by hundred and the corresponding graphs are shown as Figure~\ref{netimp}.

As a consequence of change in the network diffusion rate, we observe that if a susceptible node becomes infected the following occurs:
\begin{itemize}
\item there will be a quick decrease of susceptible nodes in the affected diffusion set.
\item a similar trend appears in the attacked interference groups where an increase (growth) is followed by a decrease (death).
\item a similar trend is observed in the removed diffusion sets. 
\end{itemize}

\begin{figure}[h]
\centering
\subfigure[Susceptible nodes.]{
\includegraphics[scale=0.20]{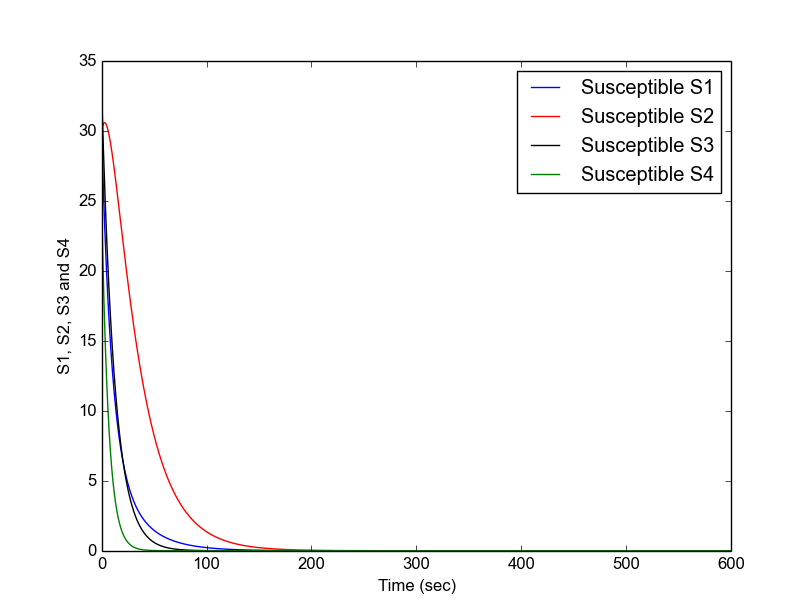} 
\label{p1}
}
\subfigure[Attacked nodes.]{
\includegraphics[scale=0.20]{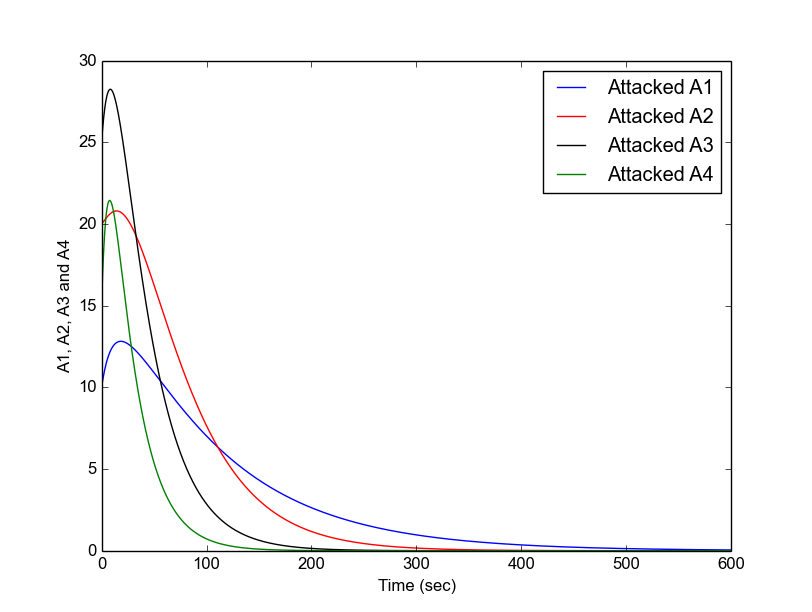} 
\label{p2}
}
\subfigure[Removed nodes.]{
\includegraphics[scale=0.20]{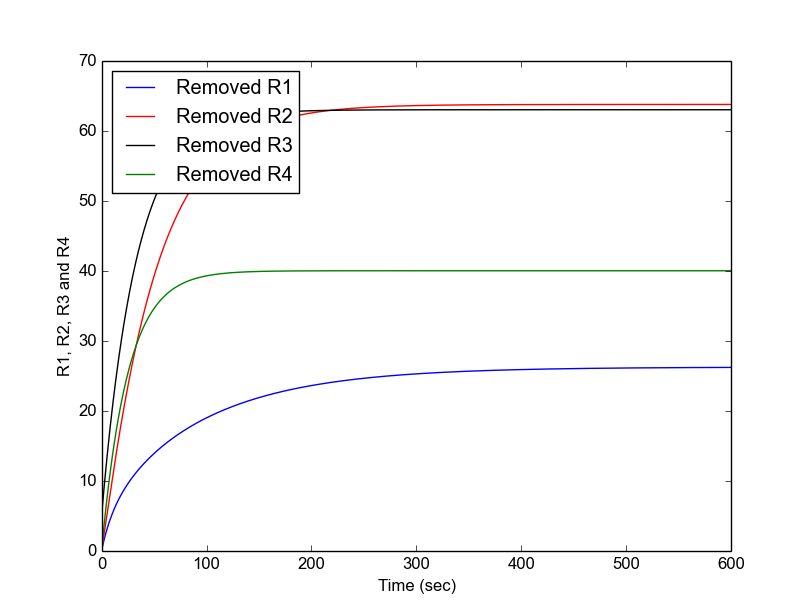} 
\label{p3}
}
\caption{Network Impact Factor: $200*\eta_2$.} \label{fig:netimp200}
\end{figure}
\begin{figure}[h]
\centering
\subfigure[Susceptible nodes.]{
\includegraphics[scale=0.20]{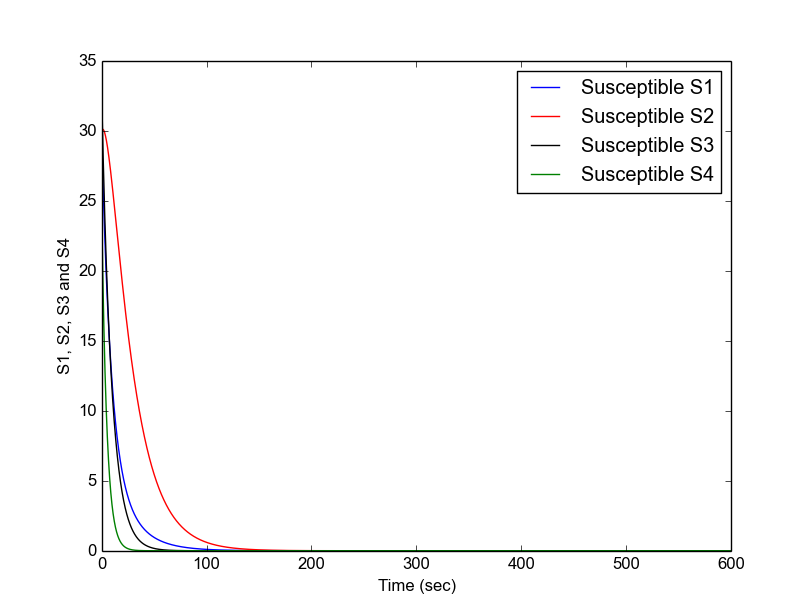} 
\label{p1}
}
\subfigure[Attacked nodes.]{
\includegraphics[scale=0.20]{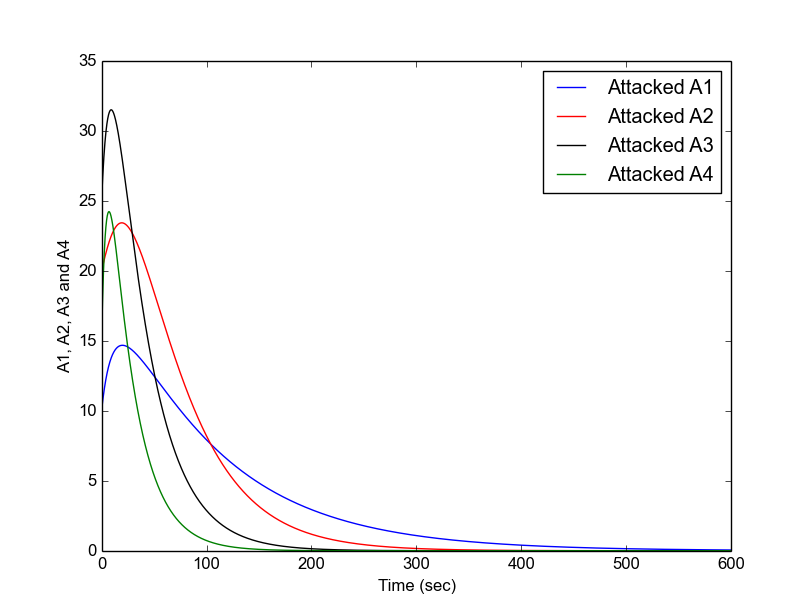} 
\label{p2}
}
\subfigure[Removed nodes.]{
\includegraphics[scale=0.20]{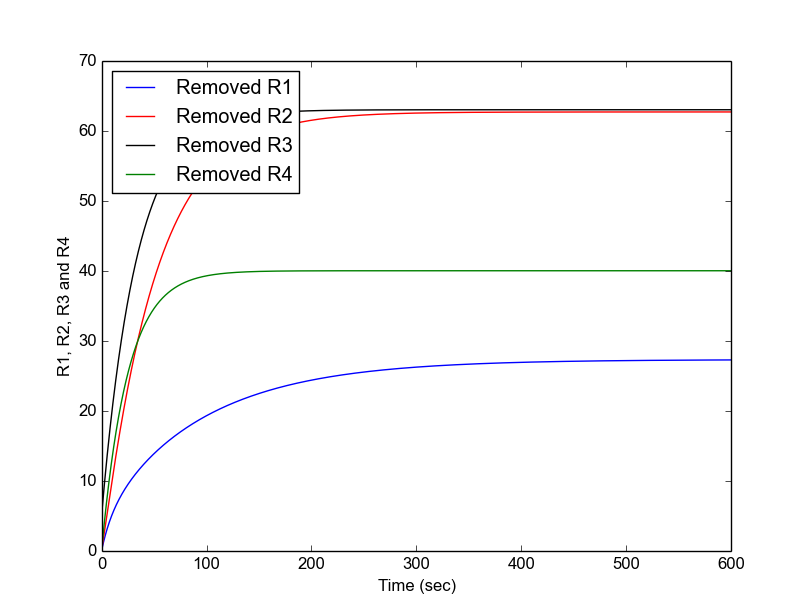} 
\label{p3}
}
\caption{Network Impact Factor: $300*\eta_2$.} \label{fig:netimp300}
\end{figure}

We conducted similar experiments by multiplying the network impact by 200 and 300 respectively. The results are reported in Figures~\ref{fig:netimp200} and~\ref{fig:netimp300} respectively and reveal similar performance patterns.

Possible triggers aiming at controlling the transmission rates and network impact parameter consist of targeting specific interference sets by artificially amending their nodal weights/interferences to either delay or accelerate movement into different interference set(s). 

\section*{Deployment Considerations}\label{sec:6}

In this section, we present the economic impacts of our model on two types of real networks - social networks (such as Facebook, Skype) and a Public safety communication networks  from a user's perspective.

Facebook is a social network where the impacts of having a node affected by a virus and thus failing is less critical than having a Public safety communication network node being infected or failing, since the latter can have consequences on the lives of citizens. Similarly, though less costly than traditional telecommunication systems, a Skype node attack by a virus has less impact on humans than an attack on a Public safety communication network.  

\subsection*{Use cases} \label{data}
Public safety deals with the manipulation of sensitive information, such as fire alarms or emergency alerts, which when infected can mislead people resulting in damages, injuries and/or deaths. Therefore, the unit cost of infection (attacked status) in a Public safety 
communication network is much higher than the unit cost of susceptible and removed status. Building upon the assumptions made in Section \ref{idea}, we summarise the economic impacts of our model  (\ref{mod}) in Tables \ref{eco1}, \ref{eco2}, \ref{eco3} 
and \ref{eco4} where each cell contains the  cost  $c_{ij}$ in South African $Rands$ of a node at state $i$ in a network $j$. Tables \ref{eco1}, \ref{eco2}, \ref{eco3} and \ref{eco4} show the summarised cost per node in Facebook, Skype and Public safety communication networks.
\begin{table}[h!]
\centering
\begin{tabular}
{ l  c c c } \hline 
\textbf{States } & \textbf{Facebook} (R) & \textbf{Skype} (R)   & \textbf{Public safety} (R)  \\ \hline
Susceptible &  $5$ &  $35$ &  $110$  \\ \hline
Attacked &  $55$ &  $80$ &  $210$ \\ \hline
Removed &  $55$ &  $55$ &  $70$ \\ \hline
\end{tabular}
\caption{Economic impact for interference set 1} \label{eco1}
\end{table}

\begin{table}[h]
\centering
\begin{tabular}
{ l  c c c} \hline 
\textbf{States}  & \textbf{Facebook} (R)  & \textbf{Skype} (R)   & \textbf{Public safety} (R)  \\ \hline
Susceptible &   $10$ &   $30$ &   $105$  \\ \hline
Attacked &   $65$ &   $55$ &   $255$ \\ \hline
Removed &   $0$ &   $0$ &   $0$ \\ \hline
\end{tabular}
\caption{Economic impact for interference set 2} \label{eco2}
\end{table}

\begin{table}[h!]
\centering
\begin{tabular}
{ l  c c c } \hline 
\textbf{States } & \textbf{Facebook} (R) & \textbf{Skype} (R)   & \textbf{Public safety} (R)  \\ \hline
Susceptible &  $20$ &  $30$ &  $70$  \\ \hline
Attacked &  $40$ &  $10$ &  $150$ \\ \hline
Removed &  $50$ &  $20$ &  $40$ \\ \hline
\end{tabular}
\caption{Economic impact for interference set 3} \label{eco3}
\end{table}

\begin{table}[h!]
\centering
\begin{tabular}
{ l  c c c} \hline 
\textbf{States}  & \textbf{Facebook} (R)  & \textbf{Skype} (R)   & \textbf{Public safety} (R)  \\ \hline
Susceptible &   $20$ &   $10$ &   $50$  \\ \hline
Attacked &   $0$ &   $0$ &   $0$ \\ \hline
Removed &   $110$ &   $100$ &   $70$ \\ \hline
\end{tabular}
\caption{Economic impact for interference set 4} \label{eco4}
\end{table}

\subsection*{Economic results} 
\begin{figure}[h]
\centering
\subfigure[Average cost of susceptible nodes]{
\includegraphics[scale=0.20]{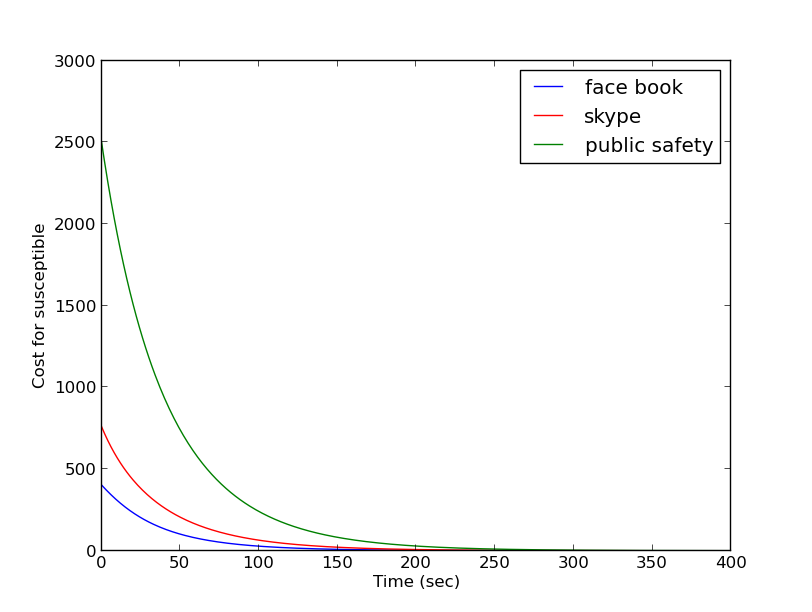} 
\label{a1}
}
\subfigure[Average cost of attacked nodes.]{
\includegraphics[scale=0.20]{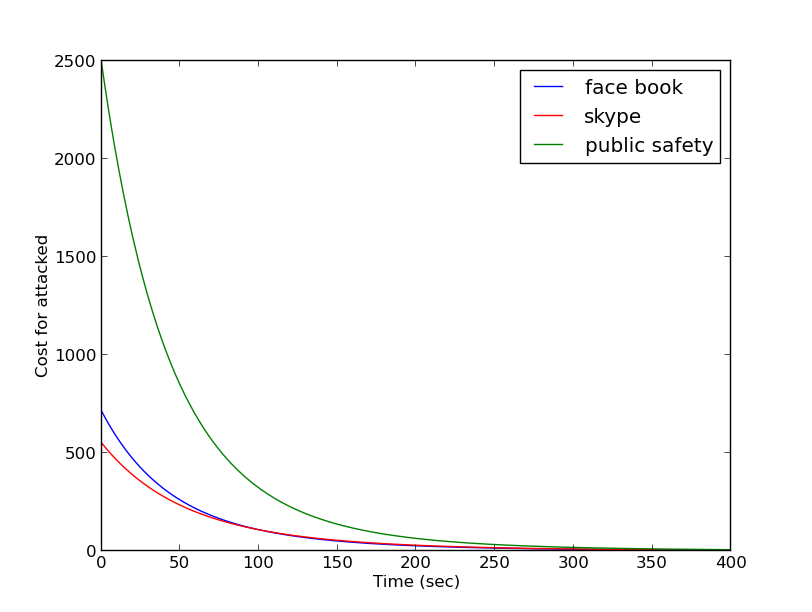} 
\label{a2}
}
\subfigure[Average cost of removed nodes.]{
\includegraphics[scale=0.20]{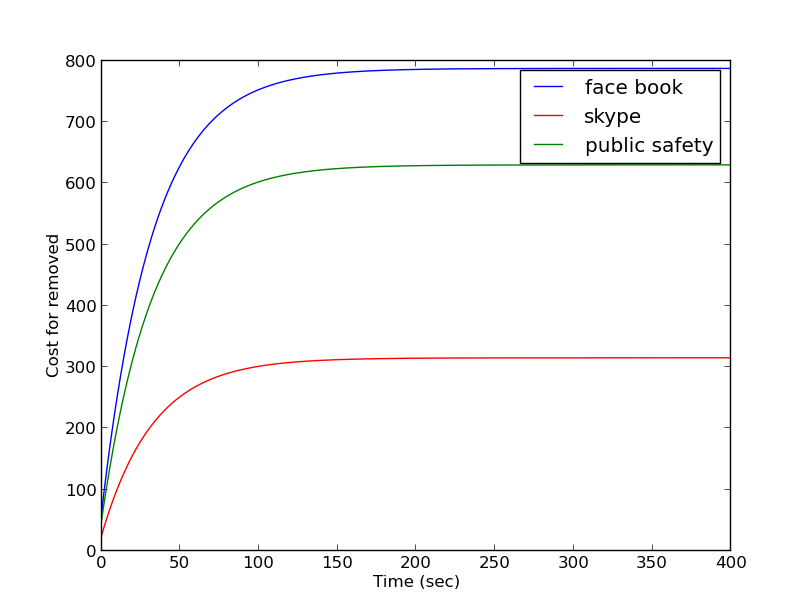} 
\label{a3}
}
\caption{Average cost of infection.}
\end{figure}

The epidemic model proposed in this paper was applied to the three types of networks described above and the results were correlated to the Tables 3 - 6 to reflect realistic costs associated with the different states of nodes within these networks. Figures \ref{a1}, \ref{a2} and \ref{a3}  reveal  the trends of the average costs for different networks and states.

Figure \ref{a1} shows that the expected average cost in public safety communication networks is always the most expensive and that all cost eventually tends towards zero. On the other hand the average cost in Facebook network is least until the  the cost in all networks tends to zero. Figure \ref{a2} shows the same trends but the average cost in Skype network is expected to remain less than that in Facebook. Figure \ref{a3} shows that the cost of removed nodes tends to a non zero value for all networks.

\section*{Conclusion} \label{sec:7}

Building upon epidemic and diffusion sets, this paper has expanded the work done in~\cite{1,2} to propose a compartmental network framework and a diffusion model which uses the SAR epidemic model to find propagation patterns between the diffusion sets and their impact on the IoT subsystem performance.  The results of the numerical evaluation of the diffusion model reveal the following: 
\begin{itemize}
\item The diffusion sets and transmission between these sets can indeed be used to quantify energy usage in wireless sensor networks, such that with increased interference, the nodes in the diffusion sets move from susceptible to attacked states and end up in the removed states where the whole network might be destroyed (energy depleted) if deployed unmaintained as mimicked by the SAR epidemic model proposed in this paper.  Furthermore, three different trends are revealed by the diffusion patterns for the evolution over time of nodes in the susceptible, attacked and removed status, which are ``birth-growth-
decrease'', ``birth-decrease'' and ``growth-plateau'' respectively. 
\item As proposed in this paper, the diffusion model uses parameters such as migration rates and network impact which can be adjusted in a real world 
deployment scenario to align a network behaviour to the service provider needs and application constraints. Such parameters could be adjusted for example to mimic an efficiently engineered sensor network, where the sensor nodes spend more time in the susceptible state and less in the attacked state before removal (death) or an economically efficient sensor network, where it is acceptable that the sensor motes spend less time in the susceptible state but stay longer in the attacked state before being removed. 
\item  The impact of the diffusion sets on the interference diffusion model is revealed by the fact that the amount of nodes in the removed class of a specific diffusion set does not necessarily amount to the initial amount of nodes in the susceptible and attacked classes of that diffusion set. This is due to the fact that, as suggested earlier, the removal/death of nodes in our model may lead to moving some nodes from one interference/epidemic set to another or the creation of new interference/epidemic sets.   
\item Though deployed in an unmaintained / autonomous mode, with predictable migration time, the proposed model is able to mitigate the destructive effect of interference on the whole network; which can hinder network management processes.
\end{itemize}

As presented in this paper, the compartmental diffusion model uses global performance parameters such as the migration rates and the network impact. As suggested earlier in this paper, these constraints may be used to lead the IoT network operation into a prescribed regime that meets a specific application requirement. Amending the LIBP protocol and its underlying algorithm to track the diffusion sets occupancy in order to move the path selection towards a given application specific behaviour is an avenue for future work. However, it will require dealing with the local parameters handled by the LIBP protocol. Achieving quality of service (QoS) in sensor networks through service differentiation  as done in~\cite{1} and/or multipath routing as suggested in~\cite{qos} is a challenging issue that can also be addressed as an extension to the work presented in this paper.  

\section*{Acknowledgements}

The authors would like to acknowledge the contribution of Audrey Bakongisa Moswa in the discussions on the diffusion model as well as other colleagues who have freely contributed to proof-reading the paper. 

\nolinenumbers


\end{document}